\documentclass[twocolumn,rmp,aps]{revtex4}

\usepackage{dcolumn,graphicx,amsmath,amssymb,pxfonts}

\begin{document}

\title{Structure and Time-Evolution of an Internet Dating Community}

\author{Petter \surname{Holme}}
\email{holme@tp.umu.se}
\affiliation{Department of Physics, Ume{\aa} University, 
  901~87 Ume{\aa}, Sweden}

\author{Christofer R.\ \surname{Edling}}
\affiliation{Department of Sociology, Stockholm University, 106~91
  Stockholm, Sweden}

\author{Fredrik \surname{Liljeros}}
\affiliation{Department of Sociology, Stockholm University, 106~91
  Stockholm, Sweden}
\affiliation{Department of Medical
    Epidemiology and Biostatistics, Karolinska Institutet S-171 77
    Solna, Sweden}

\begin{abstract}
We present statistics for the structure and time-evolution of a
network constructed from user activity in an Internet community. The
vastness and precise time resolution of an Internet community offers
unique possibilities to monitor social network formation and
dynamics. Time evolution of well-known quantities, such as clustering,
mixing (degree-degree correlations), average geodesic length, degree,
and reciprocity is studied. In contrast to earlier analyses of
scientific collaboration networks, mixing by degree between vertices
is found to be disassortative. Furthermore, both the evolutionary
trajectories of the average geodesic length and of the clustering
coefficients are found to have minima.
\end{abstract}

\maketitle

\footnotesize

We thank Christian Wollter and Michael Lokner at pussokram.com, Stefan
Praszalowicz at nioki.com, and Niklas Angemyr and Reginald Smith for
granting and helping us getting access to data. We thank Mark Newman
for comments on assortative-mixing, and the editor and anonymous
reviewer for helpful comments. PH is partially supported by the
Swedish Research Council through contract no.\ 2002-4135. CRE is supported
by the Bank of Sweden Tercentenary Foundation. FL is supported by the
National Institute of Public Health.

\normalsize

\section{Introduction}
With the growing interest in social network analysis from the physics
community, a new research area is emerging in the intersection between
statistical physics and sociology (Albert and Barab\'{a}si 2002;
Dorogovtsev and Mendes 2002; Newman 2003). Sociologists have been
interested in network analysis for at least half a century, and with
mathematicians and statisticians they have developed a set of tools to
analyze positions, structures, and processes of social networks
(Wasserman and Faust 1994; Butts 2001). Although there are exceptions
(Fararo and Sunshine 1964; Skvoretz 1990), most sociological and
anthropological studies of networks have focused on small-group
interaction or cognitive networks. In one respect this is quite
natural as most groups and formal organizations are of small
size. Also, a pragmatic reason for this is that data collection of
large social networks, behavioral or cognitive, is cumbersome and
often practically impossible to carry through. Therefore, although
recent analyses (Watts and Strogatz 1998; Watts 1999; Newman 2001)
have brought new attention to comparative analysis of large-scale
social networks, the statistical physics method, emphasizing the limit
of large system sizes (Albert and Barab\'{a}si 2002), has been of
limited utility. However, the extended use of database technology
provide new possibilities for constructing real world networks for the
analysis of e.g.\ movie-actor networks (Watts and Strogatz 1998) and
co-authorship in science (Newman 2001). Surely, these networks reflect
social interaction, but they are also heavily constrained by the logic
of a particular industry or a particular professional activity. Thus,
to allow for exploration of the possible universal properties of
social networks in general, there is still an urgent need to analyze
other types of large empirical social networks. In this paper we
report on an investigation of a large social network, aiming to give a
phenomenological description that will hopefully shed some new light
on the processes forming the structure of social networks. To put
results in context, we try to compare our findings to other studies
whenever possible, and to contrast parameters to what would be
expected from a random network with similar characteristics.

To construct network data and large graphs based on more spontaneous
patterns of human interaction than e.g.\ co-authorship and
co-actorship, one can consider data from e-mail exchange (Ebel,
Mielsch et al.\ 2002) or user activity in Internet communities
(Rothaermel and Sugiyama 2001; Smith 2002). The present work belongs
to the latter category, with a strong focus on the dynamics of the
network. In contrast to previous studies of Internet communities
(Smith 2002), we use down-to-the-second timing of the communication to
investigate time evolution and obtain steady state estimates of
well-known measures of graph structure. We use data from a Swedish
Internet community called pussokram.com (roughly ``kiss'n'hug'' in
English) that is primarily targeted at adolescents and young
adults. The community provides an arena for flirting, dating, and
other romantic communication; as well as communication for
non-romantic friendship.

Studies suggest that online interaction is driven by the same needs as
face-to-face interaction, and should not be regarded as a separate
arena but as an integrated part of modern social life (Wellman and
Haythornthwaite 2002). Thus communicative actions taken by members of
the community can be expected to share many features with the web of
human acquaintances and romances in the social off-line world. Indeed,
for many people in contemporary Western societies, interaction on the
Internet is as real as any other interaction (Wellman 2001). Internet
communities are interesting by and for themselves, but this suggests
that the formation and dynamics of social networks in an Internet
community can share the same generic properties as all social
acquaintance networks, and that the study of Internet communities can
provide important information for enhancing our understanding of
social networks in general.

The paper is divided into four sections. In the next section we give a
detailed description of the functions of the Internet community in
focus. The third section contains statistical analyses and
presentation of results that we summarize and discuss in the fourth
and concluding section.

\section{The Internet community pussokram.com\label{sec:pok}}

Pussokram.com is a Swedish Internet community primarily intended for
romantic communication and targeted at adolescents and young
adults. The community had around 30$\,$000 active users during the spring
and summer 2002, the mean user age is 21 years, and approximately 70
percent of the users are women (therefore, and to simplify, we will
use the female gender when referring to users in this paper). Both age
and sex are self reported. It is possible to have multiple accounts on
the community. A crude check on the number of accounts linked to every
unique e-mail address indicates that this is not very common (more
than 99.7\% of the membership accounts are associated with a unique
e-mail address and no e-mail address are associated with more than 5
accounts).\footnote{Of course it is possible to use an unique e-mail
  address for every unique e-mail account but since this information
  is not revealed its hard to see way on would go through the extra
  effort so doing.} Our data consists of all the user activities on
pussokram.com logged for 512 days from 13:39:25 on February 13, 2001
($t = 0$) to 13:28:19 on July 10, 2002. The smallest time-unit on the
log is 1 second. We analyze the activity of all users registered at
time $t = 0$, as well as the activity of any new users during this time
span.\footnote{Personal integrity is of course an issue here. For the
  analysis, we study the anonymized data to prevent any intrusion of
  privacy, and we do not have access to specific message
  contents. Like everyone else, we can read the guest books, but still
  we cannot link an user (and her guest book) to the vertices of the
  network. Thus, we cannot identify any specific individual person in
  the data. We do not even have data that can be cross-examined with
  other databases (like computer IP-addresses) to detect users
  identity}  Time $t = 0$ defines the start up day for this particular
community. However prior to $t = 0$ there was a mail server for
sending anonymous love messages on the Internet. Registered users of
this service had their accounts automatically transferred to
pussokram.com. We only study activity on the community, nevertheless
this recruitment might induce higher initial growth of active users.

\begin{figure*}
  \centering{\resizebox*{\linewidth}{!}{\includegraphics{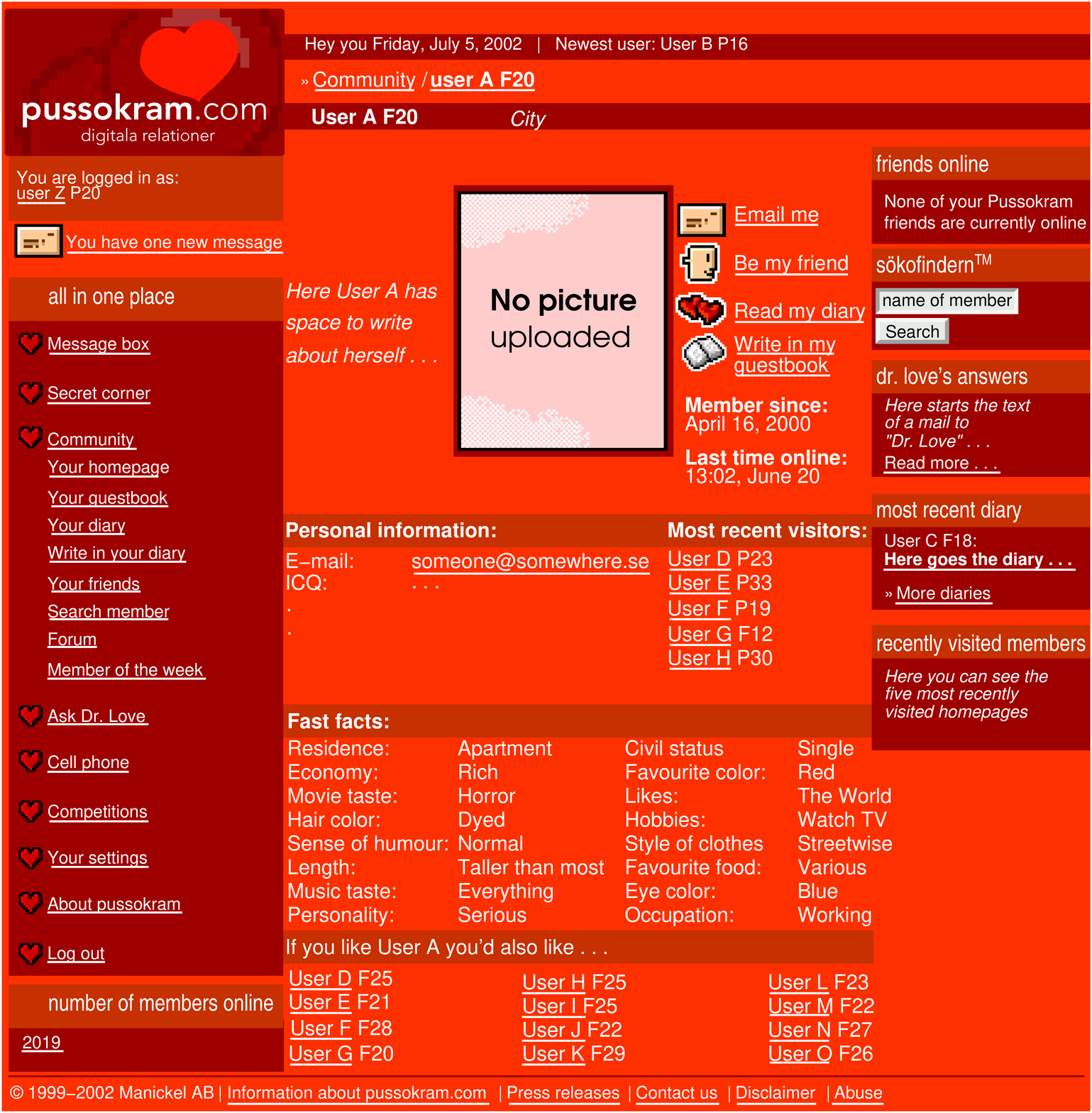}}}
  \caption{Screenshot of a typical user homepage at
    pussokram.com. ``User A'', ``User B'', etc.\ symbolize user names. (The
    translation is due to the authors. Italics denote a description
    rather than a translation.)
}
  \label{fig:pok}
\end{figure*}

Pussokram.com has a pronounced romantic profile, where:
\begin{itemize}
\item Users are encouraged to send messages to others that they are
  secretly in love with.
\item The provider answers questions related to love and sex posed by
  the users under the pseudonym Dr.\ Love.
\item The design of the HTML-pages makes use of a romantic iconography
  well known to the targeted users (with Valentine's hearts, deep red
  colors, etc., see Fig.~\ref{fig:pok}). Nevertheless, a quick glance
  through some of the public guest books reveals that many of the
  contacts taken are also non-romantic.
\end{itemize}

\subsection{Types of contacts in pussokram.com}

There are four major modes of communication at pussokram.com. We study
each of the networks generated by these four types of contacts
separately and we also study the union of these networks generated by
any of these contacts. A brief description of the four types of
contacts follows:
\begin{itemize}
\item The \textbf{Messages} are in effect intra-community e-mails. These
are private in the sense that no one in the community, except the
sender and receiver, can access them. Not even information on how many
messages other users have received are retrievable for other users.
\item In \textbf{Guest book} signing, each user has a guest book that
  every community member is free to write in.
\item \textbf{Flirt} or ``friendship request:'' User A can ask user B to
  be her friend. If user B accepts user A's request then they can both
  easily see if the other is online whenever they are logged onto
  pussokram.com. Information on the friends of a specific user is
  private to the user only.
\item \textbf{Friendship}: A friendship relation is established after
  acceptance of a friendship request, as described above. The
  friendship network is thus bi-directional. A friendship can be
  canceled by any of the friends.
\end{itemize}

\subsection{Ways to receive attention and search users}

Unless engaged in peer-to-peer contact of some sort, users at
pussokram.com are relatively anonymous towards each other. There is
reason to believe that knowledge about the prior interactive behavior
of other individuals structures the present interactive behavior of a
given individual (the so called imitation factor). The only
information about a user's interaction history available to other
users. But there are several ways for an user to draw attention to
herself (i.e.\ to direct other users to her community homepage), and
for users to find information about others. Here we summarize various
ways that can be used to receive attention, search for other users,
and promote oneself at pussokram.com. The following information is
displayed when a logged on user browse the pussokram.com website:
\begin{itemize}
\item The username of the most recently registered community member.
\item The name of the most recently edited diary (each user has space
  open for others to read, intended as a diary).
\item The names of the most recent users to browse a specific user's
  homepage.
\item The names of similar users are displayed on a specific users
  homepage. Similarity is assesses through self-reported background
  variables.
\item A long interview with the ``user of the week'' (although updated
  more seldom than weekly). This is an epithet that users can apply for.
\item Photographs of 10-20 users are displayed at the login-page.
\end{itemize}

A user can search out other users with a search engine (the
``s\"{o}kofinder''---in English ``search'n'finder''---in
Fig.~\ref{fig:pok}) that handles the following
criteria: Sub-string of the username, gender, age, place of residence,
online status, and if a user has provided a photograph of
herself. Presumably, these are the characteristics that drive user
activity, but because it is hard to assess their validity, and because
we are only interested in structural properties, we do not conduct any
analysis on them.

\subsection{Comparisons with other empirical and statistical networks}

For comparison we also use networks by instant messaging at the French
Internet community nioki.com and scientific collaboration (or, rather,
co-authorship) networks. nioki.com and pussokram.com are rather
similar, both in terms of content and design, but compared to
pussokram.com, nioki.com is even more youth oriented and not as
focused on romantic relations as pussokram.com. Besides the
possibility of searching for user names, nioki.com has two search
procedures \textit{recherche l'amiti\'{e}} (search for friendship) and
\textit{recherche l'amour} (search for love), where one can fill out
questionnaires to find other users that match ones preferences. In the
nioki.com network, an arc connects user A to user B if user B is in
user A's list of contacts (for details see (Smith 2002). In the
scientific collaboration networks (Newman 2001) the vertices are
scientists who have uploaded manuscripts to the Los Alamos preprint
repository arXiv.org, arcs are added between scientists who have
co-authored a paper. In contrast to the pussokram.com and nioki.com
networks, ties in the scientific collaboration network is
bi-directional. Note, that the pussokram.com networks are dynamic,
while we only have access to snapshot data of nioki.com and scientific
collaboration networks. For this reason we can only make comparisons
between the static properties of these networks.

In addition, following (Anderson, Butts et al.\ 1999; Pattison,
Wasserman et al.\ 2000; Shen-Orr, Milo et al.\ 2002), we compare some
observed quantities to the corresponding average values from
randomized networks with the same degree-sequence as the original. By
this approach, we examine how aspects of structures other than the
degree sequence, influences the quantities. Every known real social
network deviates from the average randomized network in a larger or
lesser extent, depending on the social forces structuring the
interaction. For example, with regards to the present case, we believe
that an Internet community network will be closer to the average
randomized network than several other types of social networks,
because time and space constraints are much less pressing than in,
e.g., a kinship network. These randomized networks are generated by
sequentially going through all directed arcs A-B, and for every such
arc randomly select another arc, C-D, and then rewire so that A-D
forms one arc, and C-B forms another. The choice of C-D is done with
uniform randomness among all arcs that would not introduce a loop or a
multiple arc. We use this algorithm to generate $\sim 3000$ networks and the
quantities are averaged over these networks. This procedure is inspired
by Roberts (2000). However it differs from Roberts in the sense that
we use sweeps over all arcs (where each arc is rewired at least once)
as the unit of iterations of the algorithm.\footnote{To be precise our
  algorithm run as follows: We go sequentially through the arc
  set $A$ (see Sect.~\ref{sec:stat}). For every arc $(v,w)$ we
  construct a set $A'$ of arcs such that if a member $(v',w')$ of $A'$
  is to be rewired with $(v,w )$---i.e.\ so that $(v,w)$ and $(v',w')$
  are replaced by $(v,w')$ and $(v',w)$---then no loops or multiple
  arcs are formed. Then we choose one of $A$'s arcs with uniform
  randomness and rewire that arc with $(v,w)$.}

\section{Statistical analysis\label{sec:stat}}

The pussokram.com network consists of all registered users and the
communication flow between these users as described
above. Communication is conceived of as directed links between
users. This is translated into a graph of vertices (users) and arcs
(ties). Vertices are added to the network the first time a registered
user is active, i.e. the first time the user sends or receives a
message, signs a guest book, or sends or accepts a friendship request
as described above. Each of these interactions defines a unique
network, and by adding an arc for any activity one gets a total
network of online activities. We thus study five networks, and for
each of them the vertex set is empty at $t = 0$. We represent the
network as a directed graph, $G = (V, A)$, where $V$ is the vertex set
and $A$ is the set of arcs, or ordered pairs of vertices. $N = |V|$
denotes the order (number of vertices) of $G$, and $M = |A|$
represents the number of arcs. Sometimes we study properties of the
undirected graph obtained by taking the reflexive closure of
$G$.\footnote{I.e.\ the graph obtained if for every $(u,v)\in A$ and
  $(v,u)\notin A$ then $(v,u)$ is added to $A$.}

\begin{figure}
  \centering{\resizebox*{\linewidth}{!}{\includegraphics{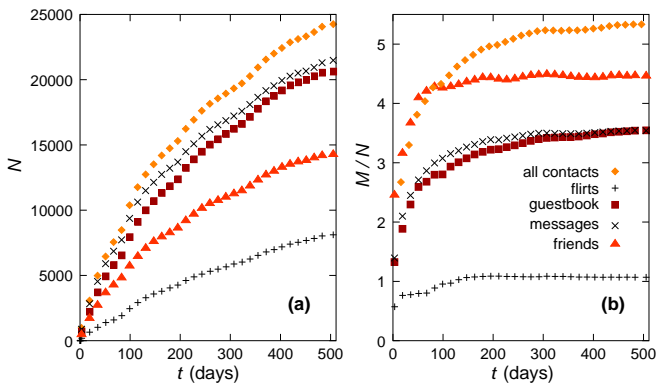}}}
  \caption{Time evolution of the number of vertices (a) and average
    degree (b) as a function of time.}
  \label{fig:len}
\end{figure}

\subsection{Decreasing growth rate of network size and convergence of
  average degree}

For each network, the number of vertices of each network, $N$, as a
function of time during the sampling is displayed in
Fig.~\ref{fig:len}(a), and the average degree, i.e.\ the average
number of arcs per vertex, $M / N$, is displayed in
Fig~\ref{fig:len}(b). As can be seen, both the number of vertices and
the average degree are increasing as a function of time, but with at a
decreasing growth rate. The average degree appears to converge to a
constant, but for $t < 100$, it increases as a power function. The
more rapid growth rate in the beginning of the period is explained by
the fact that old users log on for the first time during our sampling
period (see discussion in Section~\ref{sec:pok}). The decreasing
growth, and apparent approach to equilibrium, stand in contrast to the
accelerated growth of the Internet and the World Wide Web (Dorogovtsev
and Mendes 2002), as well the linear growth of scientific
co-authorship networks extracted from article databases (Newman 2001;
Newman 2001; Barab\'{a}si, Jeong et al.\ 2002). However, in social
networks, the average degree cannot be increasing without bounds, and
this goes for scientific collaboration networks too. We believe the
difference stems from a wider effective sampling time frame---due to
the much more rapid dynamics of an Internet community (compared to
scientific collaborations) we are, relatively speaking, able to follow
the process for a much longer period. In the sense that $G$ is a
steadily growing dynamic network, we deal with a non-equilibrium
representation of the social situation. When we speak of the network
``reaching equilibrium,'' we refer to when all quantities that are
bounded as a function of $N$ (such as the average degree) are reaching
their constant limits.

\begin{figure}
  \centering{\resizebox*{\linewidth}{!}{\includegraphics{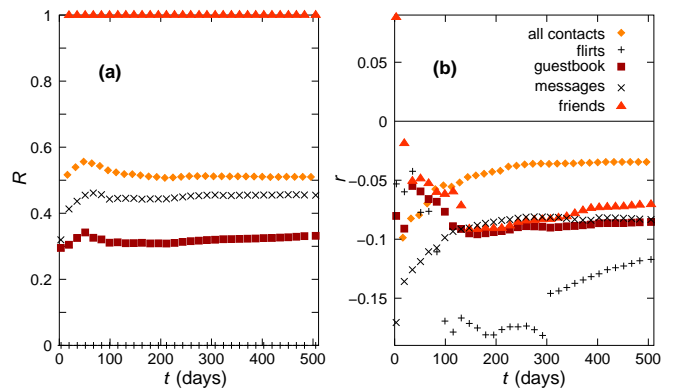}}}
  \caption{Reciprocity $R$ (a), and (b) assortative mixing coefficient
    $r_\mathrm{dir}$ as functions of time.}
  \label{fig:ass}
\end{figure}

\subsection{Reciprocity varies between networks}

Various types of social relations differ in direction, intensity, and
frequency (Granovetter 1973). Messages between agents with different
social status for example, tend to be unevenly distributed (Gould
2002). In the present analysis, we can investigate the reciprocity of
communicative action by looking at the direction of the communication
flow between any two users. For example, if user A sends a friendship
request to user B, we observe a link between user A and user B, and
note an arc between the two vertices. But it makes quite a difference
whether user B accepts the invitation or not, i.e.\ whether we note one
or two arcs between the vertices. We define reciprocity $R$, as the
fraction of mutual dyads, i.e.\ the ratio between the number of
vertex-pairs $\{v,w\}$ occur in two arcs  ($(v,w)$ and $(w,v)$) and
vertex-pairs that occur in at least one arc.  More analytically:
\begin{equation}
R=\frac{2M}{M_2}-1~.\label{eq:rec}
\end{equation}
where $M_2$ is the number of arcs in the reflexive closure of $G$. $R$
lies strictly in the interval $[0,1]$; if $(u,v)$ is an arc then $R =
0$ implies that $(v,u)$ is not an arc and $R = 1$ implies that $(v,u)$
is an arc.

\begin{table*}
\label{tab:ass}
  \caption{Assortative mixing coefficients, $r$, for five
    pussokram.com networks, and for nioki.com and arXiv.org
    networks. Statistics for corresponding randomized networks are
    within square brackets. Differences between the various mixing
    coefficients are discussed in the text. Double hyphens indicate
    missing data. Note: * $p\leq0.01$ nioki.com and arXiv.org data are
    not tested for significance.}
\begin{ruledtabular}
\begin{tabular}{l|ccccccc}
\hline
network & $N$ & $r$ & $r_\mathrm{dir}$ & $r_\mathrm{in\: in}$ &
$r_\mathrm{in\: out}$ & $r_\mathrm{out\: in}$ & $r_\mathrm{out\:
  out}$\\
all contacts & 29{$\,$}341& {--}0.048* & {--}0.059* &  {--}0.063*&{--}0.046* & {--}0.071*& {--}0.050* \\
 & & [{--}0.043]& [{--}0.041]& [{--}0.028]&[{--}0.021] & [{--}0.049]& [{--}0.035]\\
messages & 21{$\,$}545 & -­0.055* & {--}0.083*& ­0.054*& -­0.056*&  -­0.076* & -­0.087*\\
 & & [­-0.053] & [{--}0.061]&[-­0.013] &[­-0.011] & [-­0.058]& [­-0.057]\\
guest book & 20{$\,$}691 & -­0.073*& ­-0.085*&­-0.097* &­-0.043* & -­0.088*& -­0.053*\\
 & & [­-0.049] & [-­0.038]&[-­0.024] & [­-0.015]&[-­0.042] & [-­0.026]\\
friends
 & 14{$\,$}278& ­-0.042*&- - &- - & - -& - -& - -\\
 & &  [­0.031]&- - & - -& - -& - -& - -\\
flirts &8{$\,$}186 & ­-0.12*& -­0.12*& ­-0.006& ­-0.022& ­-0.12*& ­-0.042*\\
 & & [-­0.12] & [­-0.10]& [0.016]& [­-0.002]& [-­0.10]& [-­0.013]\\
nioki.com & 50{$\,$}259& -­0.13& -­0.10& -­0.088&-­0.084 & ­-0.10&-­0.095 \\
 & & [-­0.034]&[-­0.014] &[­-0.018] &[-­0.014] &[-­0.020] &[-­0.016] \\
arXiv.org &52{$\,$}909 & 0.36& - -& - -& - -&- - & - -\\
 & &  [­-0.034]& - -& - -& - -& - -& - -\\
\end{tabular}
\end{ruledtabular}
\end{table*}

The time evolution of the reciprocity can be seen in Fig.~\ref{fig:ass}a. As is
evident from the figure, reciprocity levels differ little between the
different networks. By definition, the friendship network has
reciprocity of 1. And by the same token, the flirt network has a
reciprocity equal to zero. For the other two networks, the curves
converge to values around 0.4 for the guest book and messages
networks, and 0.5 for the all contacts network (see Table~\ref{tab:ass}). 
It's hard to judge whether these are high or low values of
reciprocity. They are however compatible with data for the French
Internet community nioki.com. We normally assume acquaintance networks
to have a high degree of reciprocity, but one reason to expect a lower
value for online interaction is that an actor feels less social
pressure to respond to a communicative act over the Internet than in a
face-to-face, or telephone encounter, for example.

\subsection{Disassortative mixing coefficients of the pussokram.com networks}

Together with the degree distribution, the degree-degree correlation
is considered to govern much of the network's robustness towards
disturbances as well as the information flow. In other contexts the
discussion is usually phrased in terms of resilience against epidemics
and attack. A positive degree-degree correlation is also referred to
as assortative mixing by degree, and it means that vertices of 
high degree preferably attaches to each other, and vice versa. For
example, assortative mixing makes the networks more vulnerable to
outbreaks of diseases, and more robust against strategic attack
(Newman 2002), because if people with many contacts are connected to
other people with many contacts, the epidemic threshold will be
lowered. Disassortative mixing, on the other hand, gives rise to
larger epidemics (Morris and Kretzschmar 1995).

We measure assortative mixing by calculating Pearson's correlation
coefficient $r$ for the degrees at either side of an edge as suggested
by Newman (2002):
\begin{equation}\label{eq:r}
r=\frac{\langle k_\mathrm{to} k_\mathrm{from}\rangle -\langle
  k_\mathrm{to}\rangle\langle k_\mathrm{from}\rangle} {\sqrt{\langle
  k_\mathrm{to}^2\rangle-\langle k_\mathrm{to}\rangle^2}\sqrt{\langle
  k_\mathrm{from}^2\rangle-\langle k_\mathrm{from}\rangle^2}} 
\end{equation}

In equation \ref{eq:r},  $\langle\cdots\rangle$  denotes the average
over arcs, $k_\mathrm{from}$ is some (in-, out-, or total) degree of
the vertex that the arc starts from, and $k_\mathrm{to}$ is some degree of the
vertex that the arc leads to. We look at $r$ for total degree of both
bi-directional (where the reflexive closure has been taken if the
network is not bi-directional by definition) and directed graphs
$r_\mathrm{dir}$. Furthermore, we measure the four combinations of in-
and out degree correlations; e.g.\ the out-in correlation coefficient
indicates whether users that have many contacts (high out-degree)
prefers to communicate with those users that themselves receive
communication from many users (high in-degree).

The values for pussokram.com and other networks are displayed in
Table~\ref{tab:ass}. Interestingly enough all the pussokram.com networks,
as well as the nioki.com network display a significant disassortative
mixing for all types of degree-degree correlations. This is in
contrast to what have been measured for (scientific-, actor-, and
business-) collaboration networks (Newman 2002). To set these results
in perspective we also measure $r$ for a scientific collaboration
network, which clearly displays a positive assortative mixing
coefficient. Maybe an assortative mixing is significant only to
interaction in competitive areas, such as professional collaborations
(where only already big names are likely to be successful in
collaborating with other big names). This result relates to research
on exchange networks that claim that negative mixing is optimal when
actors are substitutable, as for example in friendship and dating
network (Cook, Emerson et al.\ 1983). In contrasts, professional
collaboration is positive because both knowledge and already
established channels for cooperation screen off potential alternative
collaborators. Another issue is the skewness of the degree
distribution. Intuitively, a large spread in the degree distribution
will increase the likelihood of observing negative mixing. And as can
be seen from the randomized networks in Table~\ref{tab:ass}, given the degree
distribution we would expect a negative mixing coefficient. However,
the observed coefficients are consistently, and significantly, higher
than expected. This strongly suggests that negative mixing arise from
this particular form of social interaction in which alters are
substitutable (Cook, Emerson et al.\ 1983). Note though, that some
network models, analyzing completely different forms of interaction,
with skewed degree distributions produce networks of zero or positive
assortative mixing (Newman 2002; Park and Newman 2003).

The six different assortative mixing coefficients of Table~\ref{tab:ass}
are all of the same sign and roughly of the same magnitude. This is
interesting since it suggests that the $r$-values is a result of other
structures (presumably the degree-sequence) rather than from the
behavior of individuals: There are no a priori reasons for
$r_\mathrm{in\: out}$ to be the same as e.g.\ $r_\mathrm{in\: in}$, as
a large $r_\mathrm{in\: out}$ means that actors that are active in the
community (have a high $k_\mathrm{out}$) tend to associate with those
who are successful in promoting themselves in the community (have a
high $k_\mathrm{in}$), while a large $r_\mathrm{in\: in}$ means that
the latter category has a preference towards each other.

Fig.~\ref{fig:ass}b shows the time development of the assortative
mixing coefficient $r_\mathrm{dir}$ (the time development of the other assortative
mixing coefficients of Table~\ref{tab:ass} is qualitatively
similar). We see that $r_\mathrm{dir}$ converges more quickly than the
average degree. This is not surprising since the correlation
coefficient is a function of the way ties are formed rather than the
size or average degree of the network. An interesting detail of
Fig.~\ref{fig:ass}b is the jump at $t\approx 300$ days in the flirt
(friendship request) network. This is due to the formation of a tie
between two of the most connected actors. (The fact that the flirt
network is by far the sparsest strengthens this effect.)

\subsection{Cumulative degree distributions are highly skewed}

\begin{figure*}
  \centering{\resizebox*{0.7\linewidth}{!}{\includegraphics{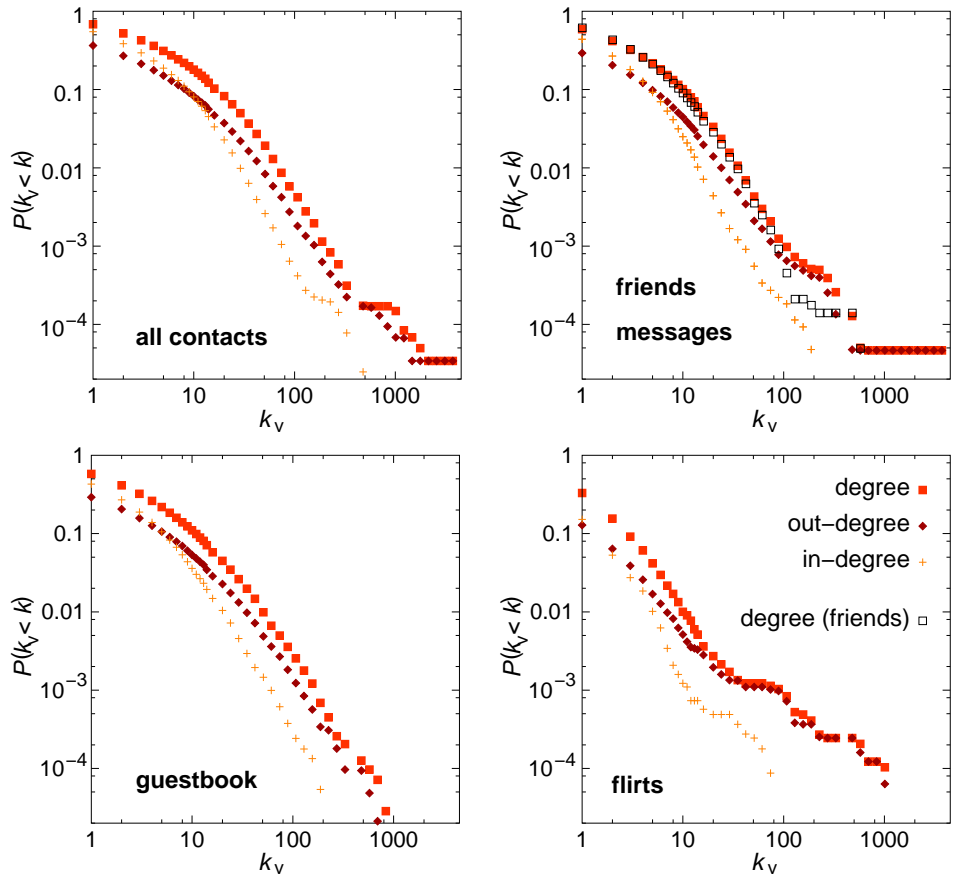}}}
  \caption{Cumulative degree distribution for the networks at the
    largest times, for all contacts (a), friendship confirmations and
    messages (b), guest book (c), and flirts (d).}
  \label{fig:deg}
\end{figure*}

The degree distribution has received much attention in comparative
analyses of complex networks since the work of Barab\'{a}si and Albert
(1999). A skewed degree distribution is commonly regarded as a
cumulative effect in the attachment of new arcs to the network (Simon
1955; Barab\'{a}si and Albert 1999), and it offers a way to classify
different types of networks (Amaral, Scala et al.\ 2000). Indeed it
has been demonstrated that many apparently dissimilar types of
networks share the same highly skewed degree distributions of a
(truncated) power-law form (Albert and Barab\'{a}si 2002), indicating
an emerging scale-free structure. Such degree distributions are
generated through a growth process in which new arcs are drawn between
already existing vertices and new vertices only. However, a process
that reasonably describes the activity of an Internet community would
allow also for new arcs to be drawn between two already existing
vertices. Such a mixed process however, would result in a stretched
exponential distribution, and not a power-law, and thus a stretched
exponential distribution is what we would expect to observe. Another
process that can be responsible for cutting the tails of power-law
degree distributions in real-world networks is a limited capacity of
the actors.

Following (Liljeros, Edling et al.\ 2001) we measure the cumulative
degree distribution of all the pussokram.com networks, see
Fig.~\ref{fig:deg}. If the degree distribution follows a power-law
with exponent $­\gamma$ then the cumulative distribution will have the
exponent $­\alpha = ­\gamma + 1$. All pussokram.com networks are
highly skewed, but none of them fits a power-law form across the whole
range observed. However, it is interesting to note that there are no
clear signs of the (inevitable) high-degree truncation in any of the
graphs (Fig.~\ref{fig:deg}). A previous study of the French nioki.com
has reported a power-law fit of the cumulative degree distribution
(Smith 2002). Our result might appear to set the pussokram.com
community apart from the nioki.com community, but a closer inspection
of our graphs and (Smith 2002) reveals a striking similarity in the
functional form of the distribution. We therefore conclude that the
dynamics shaping the degree-distribution is to a large extent the same
for the two communities.

\subsection{Evolution of average geodesic length}

As a general measure of how closely connected a graph is, the average
geodesic (shortest path) length is one of the most studied network
quantities. There is no unique natural definition of average geodesic
length in an arbitrary directed graph{--}-the problem is the
contribution from disconnected pairs of vertices. One choice is to
measure the geodesic distance averaged over pairs of vertices in the
giant component:
\begin{equation}
l_\mathrm{GC}=\frac{1}{|A_\mathrm{GC}|} \sum_{(u,v)\in A_\mathrm{GC}}d(u,v)~,
\end{equation}
where $d(u, v)$ is the distance between $u$ and $v$, and
$A_\mathrm{GC}$ is the arc-set of the giant component. Another option
is to average the inverse geodesic length (Latora and Marchiori 2001), 
\begin{equation}
l^{-1}=\frac{1}{M} \sum_{(u,v)\in A}\frac{1}{d(u,v)}~,
\end{equation}
where $1/d(u, v)$ is defined as zero when no path exists from $u$ to
$v$. In the present paper we focus on $l^{-1}$, and $l_\mathrm{GC}$
for the reflexive closure of $G$. If the two measures agree, we can
infer that there is no additional effect influencing the shortest
paths in a substantial way, other than the bi-directional structure of
the largest connected subgraph.

\begin{figure}
  \centering{\resizebox*{\linewidth}{!}{\includegraphics{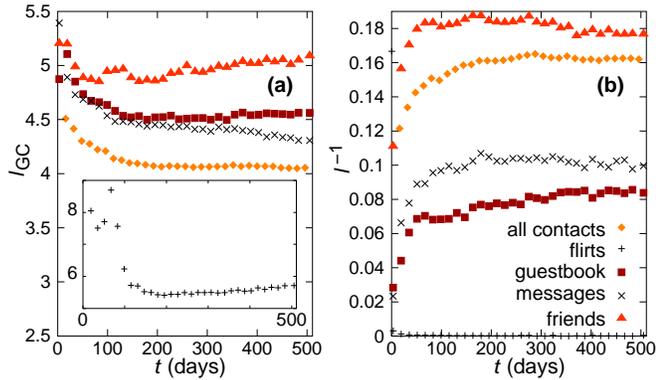}}}
  \caption{Time evolution of the average geodesic length within (a)
    the giant component of the reflexive closure and (b) the average
    inverse degree.}
  \label{fig:xlen}
\end{figure}

As time evolves there are two conflicting mechanisms governing the
average geodesic length: The increasing number of vertices works for
an increase of $l$, whereas the increasing average degree makes $l$
shorter. For the pussokram.com data the latter effect dominates,
during the time span of our data set, to give a monotonously
decreasing $l_\mathrm{GC}$ (monotonously increasing $l^{-1}$) as shown
in Fig.~\ref{fig:xlen}. The same situation has been reported for
scientific collaboration networks (Barab\'{a}si, Jeong et al.\
2002). Assuming the community outlives its members, $l$ will
eventually start to increase (when the number of inactive users slows
down the accelerated growth sufficiently).

\subsection{Density of short circuits}

Acquaintance networks are expected to have a high degree of
transitivity (Wasserman and Faust 1994), or in other words, a high
density of triangles, since if person A knows person B and person C,
then person B and person C are likely to be acquainted. We apply a
commonly used measure that gives the fraction of triangles out of the
connected 3-paths of the graph (a quantity that was defined for
undirected graphs, but is trivially generalized to directed graphs,
for which we use subscript ``dir''). If we let $p(n)$ denote the
number of representations of paths\footnote{A representation of a path
  of length three is a triplet $(u,v,w)$ such that $(u,v)$ and $(v,w)$
  are arcs. In an undirected network a path have two representations
  and a triangle has six representations.} and $c(n)$ denote the
number of representations of circuits, of length $n$, then we can
express the clustering coefficient,\footnote{This quantity is
  sometimes called transitivity, sometimes clustering
  coefficient. Note however that is not identical to Watts and
  Strogatz's (1998) clustering coefficient (where they average a local
  transitivity measure over the vertex set).} $C$, as:
\begin{equation}
C=\frac{c(3)}{p(3)}
\end{equation}
One can expect that social networks with many heterosexual romantic
relationships, such as the pussokram.com networks, to have rather few
triangles.\footnote{Presumably, homosexual relationships are not the
  common type of romantic relationship among Swedish
  adolescents. Therefore we expect few triangles. As a corollary, in a
  community populated largely by homosexual individuals, the number of
  triangles would be much higher. Regrettably we cannot test this
  hypothesis with available data.
} To get a better picture of the density of short circuits we also
measure the density of circuits of length four:
\begin{equation}
D=\frac{c(4)}{p(4)}
\end{equation}
The $n$-behavior of $c(n) / p(n)$ varies from network to network, and
could possibly be an informative quantity in it self. A very high $C$
will in most cases probably imply a high $D$ (for $R = 1$ network, two
triangles with one arc in common will contribute to $c(4)$), but the
reverse is less certain.

\begin{table*}
\label{tab:misc}
  \caption{Statistics for the fully-grown networks of
    pussokram.com, nioki.com and arXiv.org networks provided for
    comparison. Statistics for corresponding randomized networks are
    within square brackets. Double hyphens indicate missing
    data. Note: * $p\leq 0.01$. ${}^\dagger$The `friends' and
    `arXiv.org' data sets are undirected, $M$ denotes the number of
    undirected edges (which is half the number of $M$ in a directed
    representation of the graph). nioki.com and arXiv.org data are not
    tested for significance. 
}
\begin{ruledtabular}
\begin{tabular}{l|ccccccc}
\hline
network & all contacts & messages & guest book & friends & flirts & nioki.com & arXiv.org\\
$N$ & 29{$\,$}341 & 20{$\,$}691 & 21{$\,$}545 & 14{$\,$}278 & 8{$\,$}186 & 50{$\,$}259 & 52{$\,$}909 \\
$M$ & 174{$\,$}662 & 76{$\,$}257 & 73{$\,$}346 & 31{$\,$}871$^\dagger$ & 8{$\,$}744 & 405{$\,$}742 & 490{$\,$}600$^\dagger$\\
$R$ & 0.51 & 0.40 & 0.38 & 1 & 0 & 0.69 & 1\\
$l_\mathrm{GC}$ & 4.4 & 4.3 & 4.6 & 5.1 & 5.7 & 4.1 & 6.1\\
$l^{-1}$ & 0.12 & 0.10 & 0.084 & 0.18 & $4.0\times 10^{­4}$ & 0.209 &
0.121\\
$C$ & 0.006 & 0.001* & 0.014* & 0.020* & 0 & 0.0065 & 0.45\\
 & [0.006] & [0.002] & [0.007] & [0.0044] & [0.001] & [0.0081] &
[0.0020]\\
$C_\mathrm{dir}$ & 0.012* & 0.005* & 0.014* &  - - & 0* & 0.0076 & -
-\\
& [0.007] & [0.003] & [0.005] & [0] & [0.0077] & \\
$D$ & 0.017 & 0.006* & 0.022* & 0.020* & 0.212* & 0.013 & 0.35\\
& [0.009] & [0.004] & [0.008] & [0.004] & [0.004] & [0.0081] &
[0.0021]\\
$D_\mathrm{dir}$ & 0.016* & 0.008* & 0.015* & - - & 0 & 0.016 & - -\\
& [0.007] & [0.003] & [0.005] & [0] & [0.0077] & \\
\end{tabular}
\end{ruledtabular}
\end{table*}

Values for $C_\mathrm{dir}$ and $D_\mathrm{dir}$ and their undirected
counterparts are shown in Table~\ref{tab:misc}. We note that, with a
few exceptions, the values for the real networks are significantly
larger than the randomized; the difference, however, is far less
dramatic than for the scientific collaboration network. This is
contrast between the Internet community networks and the arXiv.org
data is easily explained from the fact that a paper with
$n_\mathrm{auth}\geq 3$ authors represents a fully connected subgraph of
$G$ (contributing with $n_{\mathrm{auth}}(n_{\mathrm{auth}} {-}1)
(n_{\mathrm{auth}} {-}2) / 3$ triangles). However, we would like to
stress that the values themselves are not very informative, compared
to their time dependence.

\begin{figure*}
  \centering{\resizebox*{0.65\linewidth}{!}{\includegraphics{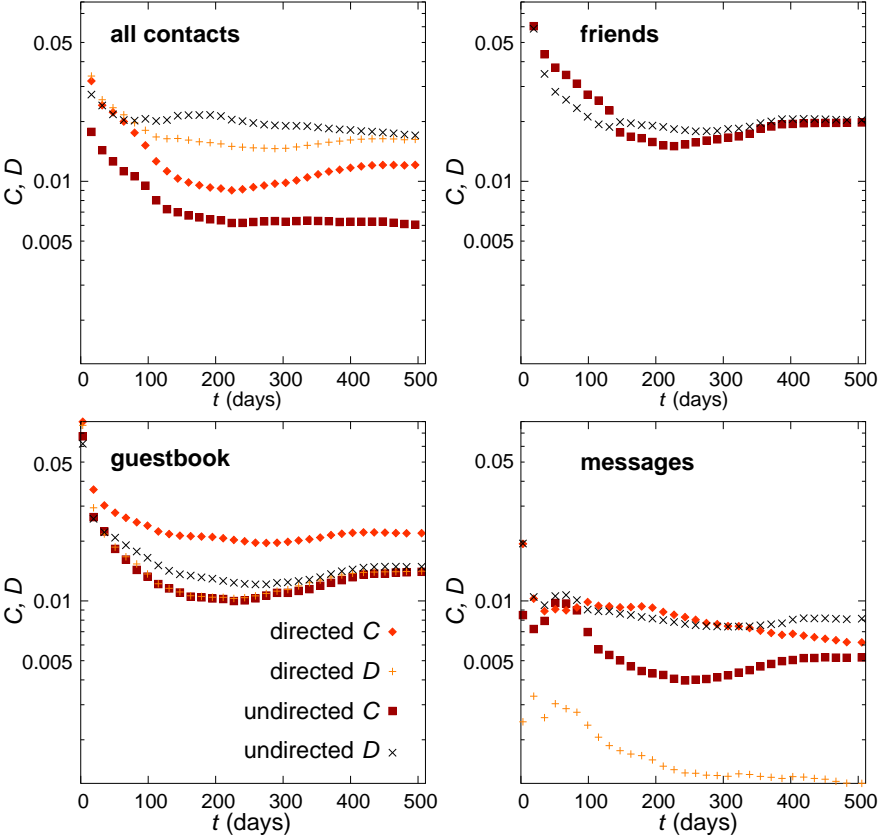}}}
  \caption{Density of short circuits for the different networks (flirt
    network omitted as it contains very few 3- and 4-circuits).
  }
  \label{fig:clu}
\end{figure*}

The time development of $C$ and $D$ for different networks is shown in
Fig.~\ref{fig:clu}. As a quantity dependent on only the local network
structure the density of short circuits is an intrinsic quantity; and,
as seen for the clustering coefficient (Barab\'{a}si, Jeong et al.\ 2002),
these quantities approach their equilibrium values from
above. Interestingly, just as for the assortative mixing coefficient,
the relaxation towards equilibrium is faster for $C$ and $D$ than for
the average degree $M / N$; i.e.\ the density of short cycles is
rather independent of the average degree.

As can be seen in Fig.~\ref{fig:clu}, most $C$ and $D$ curves have
extremes in the middle of the time range (the density of short
circuits are at their minima). The reason for this comes from a
conflict between counteracting mechanisms of different
time-scales. There are three natural time-scales in the system: The
average time between new registrations; the average time between new
contacts for an individual user; and the average life span of a user
in the community. The latter time-scale should be responsible for the
long-term behavior such as the increase towards equilibrium of $M /
N$. And as shorter circuits are more likely in a dense network, it is
natural that $C$ and $D$ increase in the large $t$ limit. The decrease
for early times is a finite size effect that can be seen in evolving
network models with constant average degree such as the
Barab\'{a}si-Albert model (Barab\'{a}si and Albert 1999; Barab\'{a}si,
Albert et al.\ 1999; Barab\'{a}si, Jeong et al.\ 2002) and extensions
(Holme and Kim 2002), where the $C$ and $D$ curves converge from
above.

Another interesting aspect is that the values of $C$ and $D$, although
finite in the large $t$ limit, is much smaller than in the actor- and
scientific-collaboration networks. In an Internet community the way by
which people introduce strangers among their acquaintances to each
other (Newman 2001; Holme and Kim 2002) is likely not the mechanism
responsible for the finite clustering (remember that in network models
such as the Erd\"{o}s-R\'{e}nyi (1959) and Barab\'{a}si-Albert
(Barab\'{a}si and Albert 1999; Barab\'{a}si, Albert et al.\ 1999;
Barab\'{a}si, Jeong et al.\ 2002) models the clustering goes to zero as
the network grows). Instead a finite density of short circuits can be
explained by the tendency formulated in the proverbial
like-attracts-like, where the similarity is defined by signaled
social, psychological, and physiological traits.\footnote{Another
  possible explanation for the convergence of $C$ and $D$ to finite
  values is that short circuits are introduced from the offline world
  outside the community. Reading users' guest books, however, gives
  the impression that the vast majority of community-dyads were
  strangers offline. We believe that this effect is negligible, but we
  are unfortunately unable to go beyond speculation on this point.}

\begin{figure}
  \centering{\resizebox*{\linewidth}{!}{\includegraphics{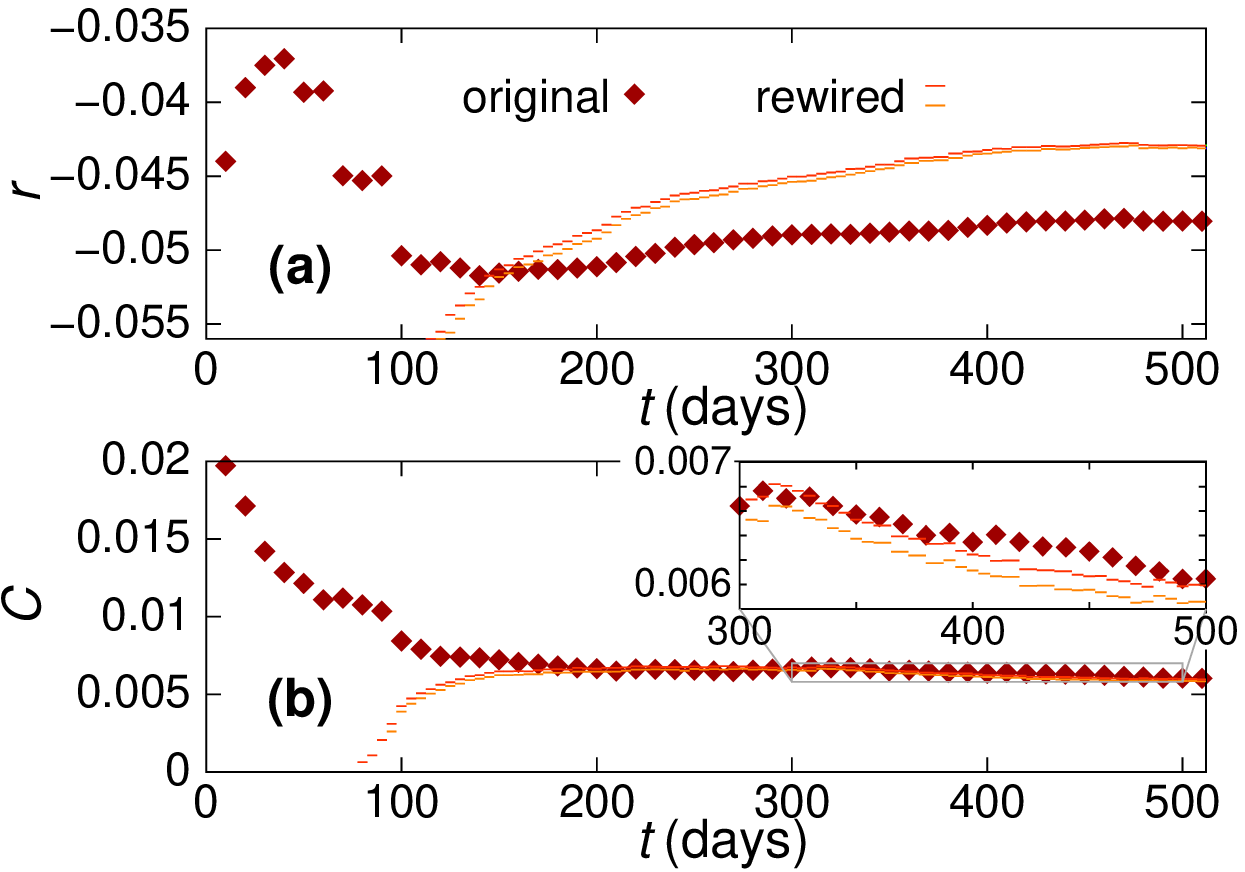}}}
  \caption{Time evolution of original and rewired quantities. (a)
    shows data for the assortative mixing coefficient $r$ for the
    undirected all-contacts network, (b) is the clustering coefficient
    for the same data. The rewired data is obtained from 100 updating
    sweeps over all links, and indicated by the upper and lower hinges
    (border values between the first and second quartile, and third
    and fourth quartile respectively).}
  \label{fig:rewi}
\end{figure}

To further convince ourselves that the sampling time is large enough
we also use rewiring to examine the time evolution of two structural
measures (the assortative mixing coefficient and the clustering
coefficient for the undirected all-contacts network). As seen in
Fig.~\ref{fig:rewi} the rewired quantities converge in the same time
scale as $r$ and $C$, which reconfirm that the sampling time frame is
sufficient. We note that for $k > 200$ days the assortative mixing
coefficient is significantly lower than the rewired reference
curve. For the same time interval the rewired clustering coefficient closely overlap the measured $C$-value; for $t > 200$ days the actual value overlap the
mid-quartiles of the rewired data during around 30\% of the 512
days. For the initial `non-equilibrium' part ($t < 100$ days) of the
time-evolution the curves of the rewired and real networks
diverges. In this region the network is rather sparse (see
Fig~\ref{fig:rewi}) which explains the low $C$-values for the rewired
$C$-curve. The high early values of $C$ seems contradictory to the
apparent absence of tendency towards triangle formation during latter
times. This means that the contact patterns of the early network is no
the same as later on. As it turns out, in the early community, a group
of actors contact each other rather frequently (rather more like
`chatting' than romantic contact making) whereas another group makes a
few contacts before quitting the community. We interpret this such
that it requires a minimal number, or ``critical mass'' (cf.\
Schelling 1978) of people for the community to function. Before the
critical mass is reached, the users either have the community as a
chat room (a usage with a presumably smaller critical mass) or leave
it.

\section{Summary and conclusions}

We have investigated networks of communication between the users of
the Internet community pussokram.com. The four different means of
contact at pussokram.com defines five different networks in our study
(one for each separately and one for all taken together). Apart from
recent studies of scientific collaboration networks and movie actor
networks, there are very few such phenomenological descriptions of
large social networks, and thus there is limited knowledge that our
findings can be related to.

It is obvious that the fact that the interaction under study takes
place on the Internet creates special conditions for communication. We
believe that the interaction online is exposed to less structural
forces than what is typically the case in most other social
settings. For example, simultaneous interaction is not a prerequisite
for communication in an Internet community, i.e.\ time as a structural
force is therefore of less importance than in most other
settings. Neither does geographical space constraint
communication. And in addition, that social signifiers are less
visible (compared to e.g.\ face-to-face interaction), and the relative
ease with which you can conceal your identity and transform your
appearance in online interaction, are factors reducing the structure
forming forces at work in `offline' social activity. It is therefore
interesting to note, that despite these caveats, the networks under
study here are much more structured than what would be expected in a
random network.

To summarize our findings of the Internet community pussokram.com, we
see that:
\begin{itemize}
\item The average degree converges over time, but surprisingly we
    observe no cut-off in the degree distribution. Previous studies do
    suggest that there is an upper limit to the mean number of
    contacts (Marsden 1987), and on average we find this
    socio-cognitive limitation despite the fact that time and space is
    of less important here. The reason we see continued growth in the
    cumulative degree distribution might be that it's relatively
    costless to have a high turnover on ones contacts in an online
    community. Contacts are established without much investment, and
    can also be dropped without much sanctioning. 
\item Reciprocity is rather low, and presumably lower can be expected
  in a regular acquaintance network. Reciprocity levels quickly
  converge to a steady state. 
\item Most assortative mixing coefficients have small negative values,
  suggesting a pattern of dissasortative mixing. This can partly be
  explained by the conventional effect from the skewed degree sequence
  (Newman 2002). The observed effect is significantly larger than can
  be expected solely from the degree distribution. An explanation for
  these higher $r$-values is the particular nature of the dating
  interaction (Cook, Emerson et al.\ 1983). We also find that mixing
  coefficients as a function of time converge rapidly. The
  dissasortative mixing in the Internet community networks is in
  striking contrast to the strong assortative mixing seen in
  scientific collaboration networks, and the nice correspondence with
  previous work in sociology indicates that Internet communities
  indeed strongly resembles off-line social communities.
\item The cumulative degree distributions are highly skewed, being a
  mixture of previous mappings of acquaintance networks (Amaral, Scala
  et al.\ 2000)---for few contacts---and partnership networks
  (Liljeros, Edling et al.\ 2001)---for many contacts.
\item The geodesic length initially increases as new vertices are
 added to the network. But as the network settles the increase is
 limited by the growing average degree. Both $l_\mathrm{GC}$ and
 $l^{-1}$ shows consistently that the average geodesic length is
 decreasing during the whole sample period (a situation that can only
 exist for a non-equilibrium network).
\item Clustering---the density of triangles---converges over time to
  non-zero values (as opposed to completely random networks). Still,
  values are probably on a much lower level than would be expected in
  offline acquaintance networks. The explanation for these low values
  is twofold---the lack of introduction as a mechanism for
  tie-formation, and the romantic profile of pussokram.com promoting
  romantic contacts. The latter aspect is also manifested in that the
  density of 4-circuits is larger than the density of triangles for
  the pussokram.com networks. Once again, the Internet community
  networks are different from the scientific collaboration network
  where clustering is larger than the density of 4-circuits.
\end{itemize}
An Internet community such as pussokram.com defines a structured
social network that share more of the structuring forces with general
acquaintance networks than networks of professional collaborations
do. We believe that the precise timing resolution and fast dynamics
(giving a wide effective sampling time-frame) will make Internet
communities an invaluable object for future social networks studies of
the largest scale.

\section*{References}

"e-print arXiv:" refers (yet unpublished) to manuscripts uploaded to
the database arXiv.org.

\begin{list}{}{\setlength{\leftmargin}{5mm}\setlength{\rightmargin}{0mm}
    \setlength{\labelsep}{5mm}\setlength{\parsep}{2mm}
    \setlength{\itemindent}{-5mm}
    \setlength{\listparindent}{0mm}\setlength{\labelwidth}{0mm}
    \setlength{\itemsep}{0mm}\setlength{\partopsep}{0mm}}

\item Albert, R.\ and A.\ L.\ Barab\'{a}si (2002).\ ``Statistical
  mechanics of complex networks." \textit{Review of Modern Physics} \textbf{74}: 47-97.
\item Amaral, L.\ A.\ N., A.\ Scala, et al.\ (2000). ``Classes of
  small-world networks.'' \textit{Proceedings of the National Academy of
  Sciences of the United States of America}
  \textbf{97}(21): 11149-11152.
\item Anderson, B.\ S., C.\ Butts, et al.\ (1999). ``The interaction
  of size and density with graph-level indices.'' \textit{Social Networks}
  \textbf{21}(3): 239-267. 
\item Barab\'{a}si, A.\ L.\ and R.\ Albert (1999). ``Emergence of
  scaling in random networks.'' \textit{Science} \textbf{286}(5439): 509-512. 
\item Barab\'{a}si, A.\ L., R.\ Albert, et al.\ (1999). ``Mean-field
  theory for scale-free random networks.'' \textit{Physica A} \textbf{272}(1-2):
  173-187. 
\item Barab\'{a}si, A.\ L., H.\ Jeong, et al.\ (2002). ``Evolution of
  the social network of scientific collaborations.'' \textit{Physica A} \textbf{299}:
  559-564.
\item Butts, C.\ T.\ (2001). ``The complexity of social networks:
  theoretical and empirical findings.'' \textit{Social Networks} \textbf{23}(1): 31-71.
\item Cook, K.\ S., R.\ M.\ Emerson, et al.\ (1983). ``The
  distribution of power in exchange networks: Theory and experimental
  results.'' \textit{American Journal of} \textit{Sociology} \textbf{89}(2): 275-305. 
\item Dorogovtsev, S.\ N.\ and J.\ F.\ F.\ Mendes (2002). Accelerated
  growth of networks. Handbook of Graphs and Networks: From the Genome
  to the Internet.\ S.\ Bornholdt and H.\ G. Schuster.\ Berlin, Wiley-VCH. 
\item Dorogovtsev, S.\ N.\ and J.\ F.\ F.\ Mendes (2002). ``Evolution
  of networks.'' \textit{Advances in Physics} \textbf{51}(4): 1079-1187. 
\item Ebel, H., L.\ I.\ Mielsch, et al.\ (2002). ``Scale-free topology
  of e-mail networks.'' \textit{Physical Review E} \textbf{66}, art.\ no.\ 035103. 
\item Erd\"{o}s, P.\ and A.\ R\'{e}nyi (1959). ``On random graphs.''
  \textit{Publicationes Matematicae Debrecen} \textbf{6}: 290-297. 
\item Fararo, T.\ J.\ and M.\ H.\ Sunshine (1964). A study of a biased
  friendship net. Youth Development Center, Syracuse University,
  Syracuse. 
\item Gould, R.\ V.\ (2002). ``The origins of status hierachies: A
  formal theory and empirical test.'' \textit{American Journal of Sociology}
  \textbf{107}(5): 1143-1178. 
\item Granovetter, M.\ (1973). ``Strength of weak ties.'' \textit{American
  Journal of Sociology} \textbf{78}(6): 1360-1380. 
\item Holme, P.\ and B.\ J.\ Kim (2002). ``Growing scale-free networks
  with tunable clustering.'' \textit{Physical Review E} \textbf{65}(2): art.\ no.\ 026107.
\item Latora, V.\ and M.\ Marchiori (2001). ``Efficient behavior of
  small-world networks.'' \textit{Physical Review Letters} \textbf{87}(19): art.\ no.\
  198701.
\item Liljeros, F., C.\ R.\ Edling, et al.\ (2001). ``The web of human
  sexual contacts.'' \textit{Nature} \textbf{411}(6840): 907-908. 
\item Marsden, P.\ V.\ (1987). ``Core discussion networks of
  Americans.'' \textit{American Sociological Review} \textbf{52}(1): 122-131. 
\item Morris, M.\ and M.\ Kretzschmar (1995). ``Concurrent
  Partnerships and Transmission Dynamics in Networks.'' \textit{Social
  Networks} \textbf{17}(3-4): 299-318. 
\item Newman, M.\ E.\ J.\ (2001). ``Clustering and preferential
  attachment in growing networks.'' \textit{Physical Review E} \textbf{64}(2): art.\
  no.\ 025102.
\item Newman, M.\ E.\ J.\ (2001). ``Scientific collaboration
  networks. I. Network construction and fundamental results.''
  \textit{Physical Review E} \textbf{64}(1): art.\ no.\ 016131.
\item Newman, M.\ E.\ J.\ (2002). ``Assortative mixing in networks.''
  \textit{Physical Review Letters} \textbf{89}, art.\ no.\ 208701.
\item Newman, M.\ E.\ J.\ (2003). ``The structure and function of
  complex networks.'' \textit{SIAM Review} \textbf{45}(2): 167-256. 
\item Park, J.\ and M.\ E.\ J.\ Newman (2003). ``Origin of degree
  correlations in the Internet and other networks.'' \textit{Physical Review E}
  \textbf{68}(2), art.\ no.\ 026112.
\item Pattison, P., S.\ Wasserman, et al.\ (2000). ``Statistical
  evaluation of algebraic constraints for social networks.'' \textit{Journal
  of Mathematical Psychology} \textbf{44}: 536-568. 
\item Roberts, J.\ M.\ (2000). ``Simple methods for simulating
  sociomatrices with given marginal totals.'' \textit{Social Networks} \textbf{22}(3):
  273-283. 
\item Rothaermel, F.\ T.\ and S.\ Sugiyama (2001). ``Virtual Internet
  communities and commercial success: individual and community-level
  theory grounded in the atypical case of TimeZone.com.'' \textit{Journal of
  Management} \textbf{27}(3): 297-312. 
\item Schelling, T.\ C.\ (1978). Micromotives and macrobehavior.\ New
  York, Norton.
\item Shen-Orr, S.\ S., R.\ Milo, et al.\ (2002). ``Network motifs in
  the transcriptional regulation network of Escherichia coli.'' \textit{Nature
  Genetics} \textbf{31}(1): 64-68. 
\item Simon, H.\ A.\ (1955). ``On a class of skew distribution
  functions.'' \textit{Biometrika} \textbf{42}: 425-440. 
\item Skvoretz, J.\ (1990). ``Biased net theory: Approximations,
  simulations, and observations.'' \textit{Social Networks} \textbf{12}(3): 217-238. 
\item Smith, R.\ (2002). ``Instant Messaging as a Scale-Free
  Network.'' eprint  arXiv:cond-mat/0206378, unpublished. 
\item Wasserman, S.\ and K.\ Faust (1994). Social network analysis:
  Methods and applications. Cambridge, Cambridge University Press. 
\item Watts, D.\ J.\ (1999). ``Networks, dynamics, and the small-world
  phenomenon.'' \textit{American Journal of Sociology} \textbf{105}(2): 493-527. 
\item Watts, D.\ J.\ and S.\ H.\ Strogatz (1998). ``Collective
  dynamics of `small-world' networks.'' \textit{Nature} \textbf{393}(6684): 440-442. 
\item Wellman, B.\ (2001). ``Computer networks as social networks.''
  \textit{Science} \textbf{293}(5537): 2031-2034. 
\item Wellman, B.\ and C.\ A.\ Haythornthwaite (2002). The Internet in
  everyday life. Oxford, Blackwell. 
\end{list}

\end{document}